\documentstyle[12pt,iopconf,graphicx]{article}

\begin{document}

\title{The Effects of General Relativity on Core Collapse Supernovae}

\author{K. R. De Nisco,\dag\ S. W. Bruenn,\dag\  and A. Mezzacappa\ddag \footnote{A.M. was supported at the Oak Ridge National
Laboratory, which is managed by Lockheed Martin Energy Research Corporation under DOE contract DE-AC05-96OR22464.  S.W.B. and A.M.
gratefully acknowledge the hospitality and support of the Institute for Theoretical Physics in Santa Barbara, which is supported
by the National Science Foundation under contract PHY94-07194.}}

\affil{\dag\ Physics Department, Florida Atlantic University, Boca Raton, FL 33431}

\affil{\ddag\ Physics Division, Oak Ridge National Laboratory, Oak Ridge, TN 37831}

\beginabstract
The effects of general relativity (GR) on the hydrodynamics and neutrino transport are examined during the critical shock
reheating phase of core collapse supernovae.  We find that core collapse computed with GR hydrodynamics results in a substantially
more compact core structure out to the shock, the shock radius at stagnation being reduced by a factor of 2.  The inflow
speed of material behind the shock is also increased by a factor of 2 throughout most of the evolution.  We have developed a code
for general relativistic multigroup flux-limited diffusion (MGFLD) in static spacetimes and compared the steady-state neutrino
distributions for selected time slices of post-bounce models with those computed with Newtonian MGFLD.  The GR transport
calculations show the expected reductions in neutrino luminosities and rms energies from redshift and curvature effects. 
Although the effects of GR on the hydrodynamics and neutrino transport seem to work against shock revival, the core
configurations are sufficiently different that no firm conclusions can be drawn, except that simulations of core collapse
supernovae using Newtonian hydrodynamics and transport are not realistic.
\endabstract

\section{Introduction}

General relativity (GR) is an essential component in the realistic modeling of core collapse supernovae because of the very
strong gravitational fields in the vicinity of the collapsed core of a star.  Hydrodynamics and neutrino transport are closely
connected in this problem, and as we will show, GR can have a profound effect on each of these, especially in the critical phase
of shock reheating.  The detection of neutrinos from supernova 1987A (Bionta \emph{et al.} 1987, Hirata \emph{et al.} 1987) and
the hope of detecting neutrino signatures from future supernovae, with next-generation detectors, is additional motivation for an
accurate general relativistic treatment of the neutrino transport in numerical simulations.

\begin{figure}[t]
\label{radii}
\includegraphics[angle=-90, scale=0.65]{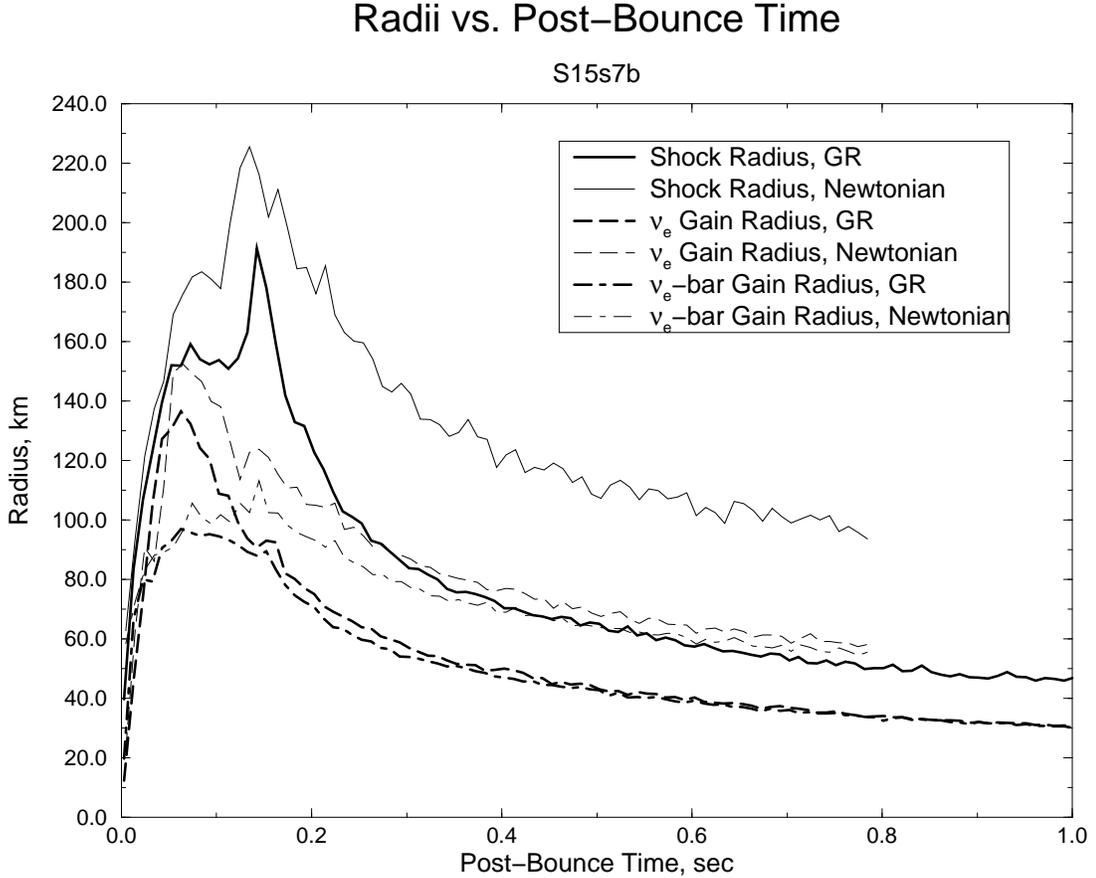}
\caption{Shock and gain radii vs. post-bounce time for model S15s7b.  Both cases are calculated with Newtonian radiation
transport.}
\end{figure}

\begin{figure}[t]
\label{rmsr}
\includegraphics[angle=-90, scale=0.65]{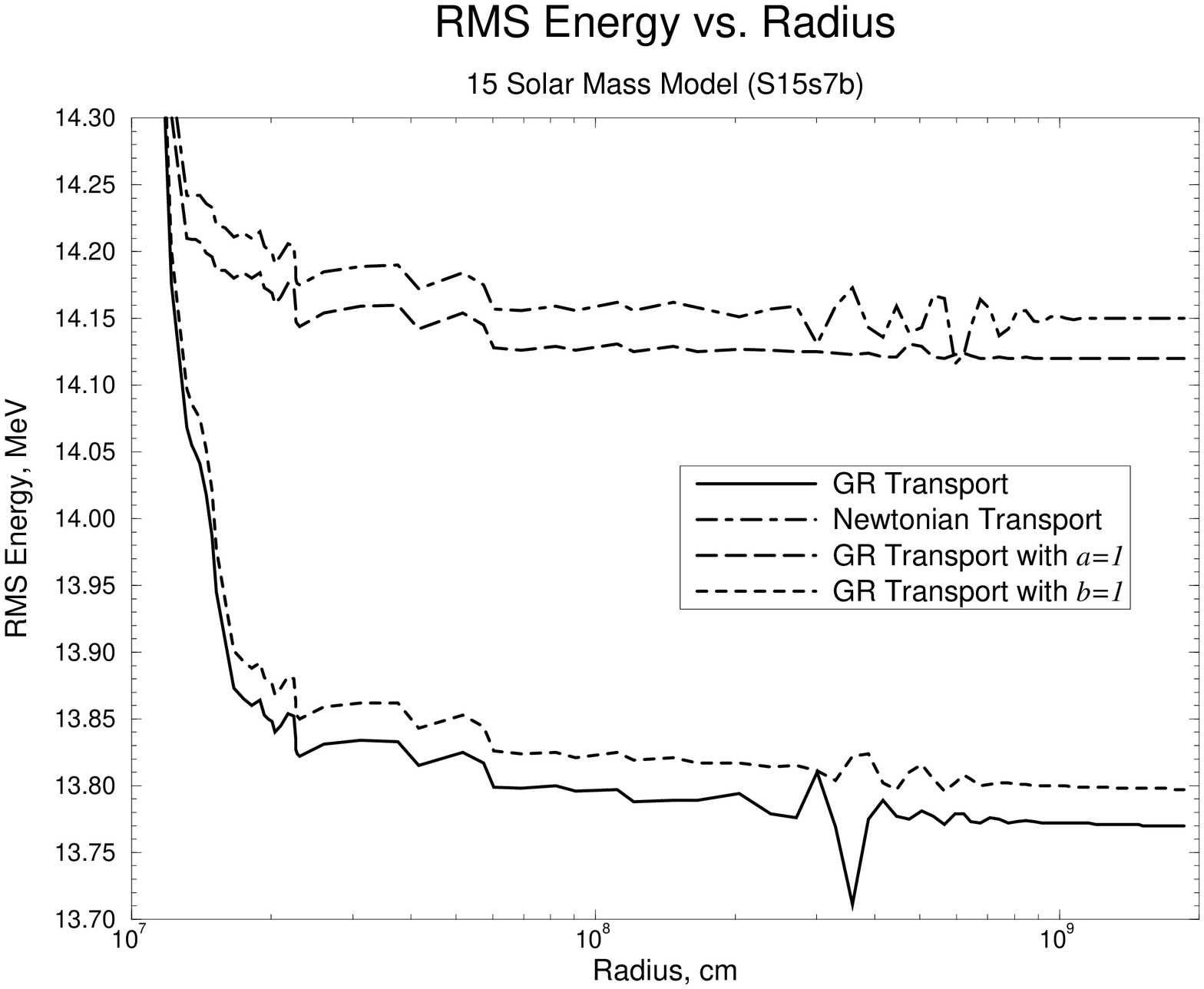}
\caption{RMS energy vs. radius for the $\nu_{\mathrm{e}}$'s in model S15s7b.  This figure contrasts stationary state GR and
Newtonian transport at $t_{\mathrm{pb}}=114$ ms.}
\end{figure}

\begin{figure}[t]
\label{lum}
\includegraphics[angle=-90, scale=0.65]{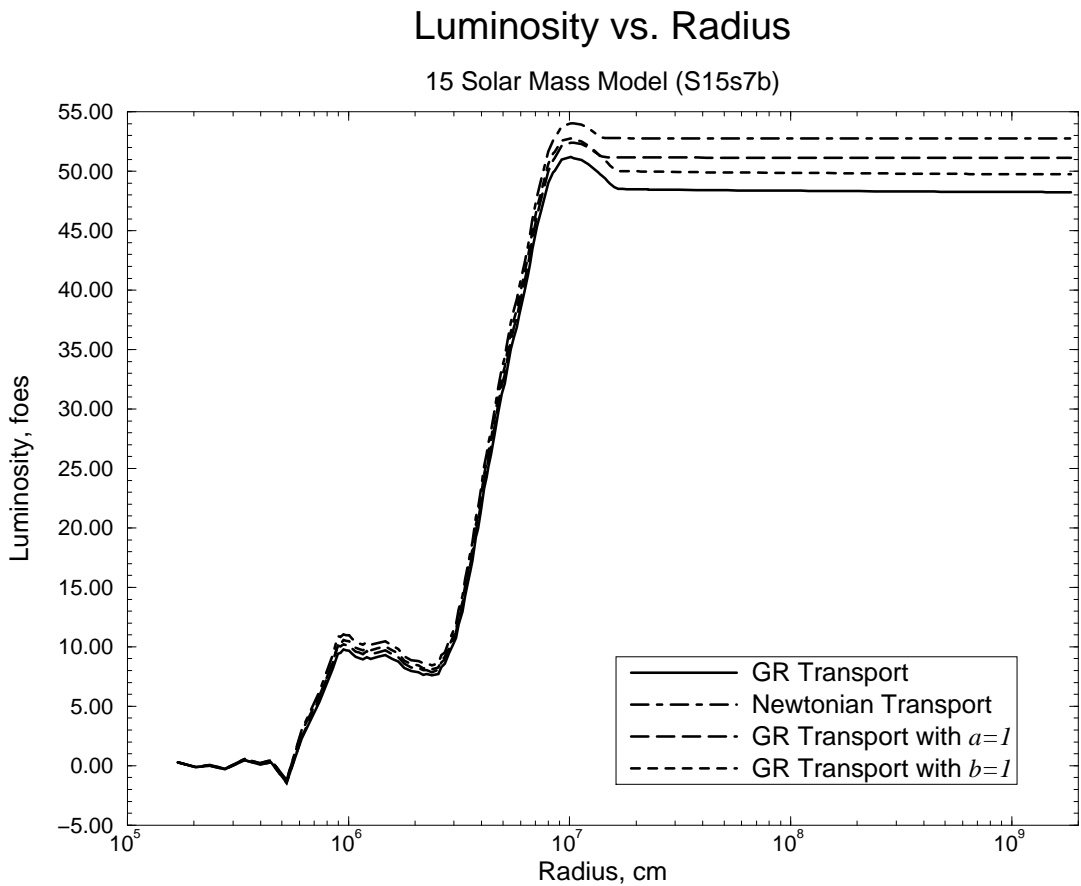}
\caption{Luminosity vs. radius for the $\nu_{\mathrm{e}}$'s in model S15s7b.  This figure contrasts stationary state GR and
Newtonian transport at
$t_{\mathrm{pb}}=114$ ms.}
\end{figure}

\begin{figure}[t]
\label{rmst}
\includegraphics[angle=-90, scale=0.65]{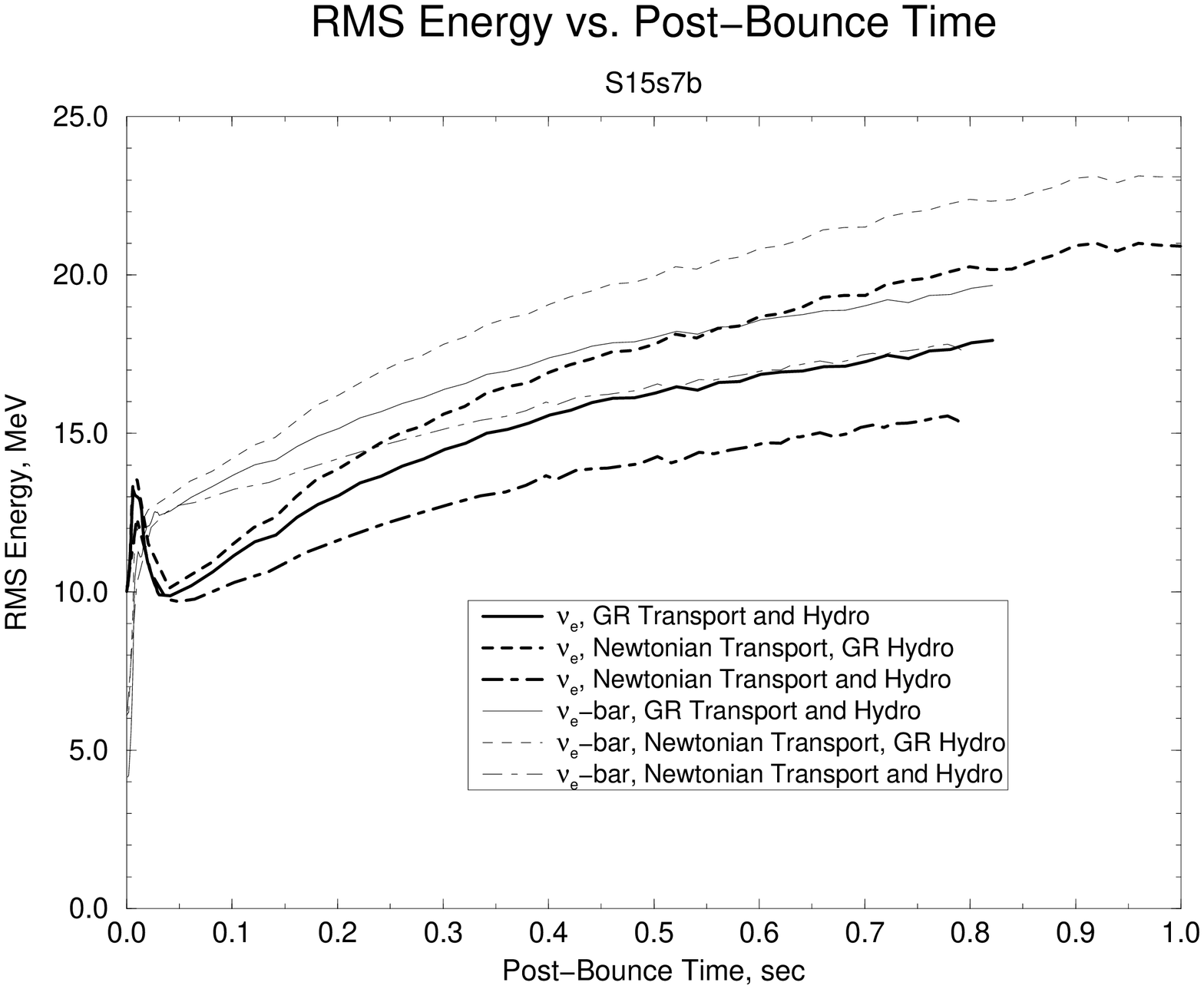}
\caption{RMS energy vs. post-bounce time for $\nu_{\mathrm{e}}$'s and $\bar{\nu}_{\mathrm{e}}$'s in model S15s7b.}
\end{figure}

We have developed a code for general relativistic multigroup flux-limited diffusion (MGFLD) that computes the neutrino transport
in a static background metric.  This is the first step in the development of a fully GR MGFLD code.  The GR metric
used is $ds^2 = a^2 c^2 dt^2 - b^2 dr^2 - r^2 (d\theta^2 + \sin^2 \theta d\phi^2)$.  This metric is allowed to evolve in the
hydrodynamics calculations, but is ``frozen'' in the transport calculations, which are then performed treating the metric as
static.  Two precollapse models, a 15$\mathrm{M}_{\odot}$ model (S15s7b) and a 25$\mathrm{M}_{\odot}$ model (S25s7b) (Woosley
\& Weaver 1995; Weaver \& Woosley 1997), were evolved through core collapse, bounce, and to approximately 800 ms after bounce in
three sets of simulations.  The first was with Newtonian hydrodynamics and Newtonian radiation transport, the second was with GR
hydrodynamics and Newtonian transport, and the third was with both GR hydrodynamics and transport.  In addition to these
three sets of simulations, stationary-state neutrino distributions were computed for various post-bounce time slices in these
models in order to isolate the effects associated with each of the metric components.

\section{The Role of General Relativity}

The effects of GR are seen quite clearly in the hydrodynamic evolution of the initial models.  GR hydrodynamics produces a much
more compact post-bounce structure than Newtonian hydrodynamics.  After $t_{\mathrm{pb}} = 0.4$ s in both models ($t_{\mathrm{pb}}$ being the
post-bounce time),  the radius of the shock and the gain radii are reduced by a factor of 2 in the GR calculations, as shown in
Figure 1.

Also strongly affected by GR is the flow velocity between the shock and the proto-neutron star.  Because matter falls through a
greater potential well to reach the shock in the GR calculation, GR preshock and therefore postshock velocities are larger than
their Newtonian counterparts, again by a factor of approximately 2.

The main effect of GR on the neutrino rms energies is the redshift of the neutrinos after they decouple from the matter, which
is governed by the metric parameter $a$.  The rms energies are reduced by a factor of $a$ evaluated at the $\nu$-sphere.  A
smaller effect is a slight outward shift of the $\nu$-sphere resulting from the (non-unity) value of the metric parameter
$b^{-1}$, which causes the neutrinos to decouple outside the $\nu$-sphere at a lower temperature.  These effects are shown in
Figure 2.  Also shown are the independent effects of $a$ and $b$ on the neutrino transport.

GR reduces the neutrino luminosity by three effects:  redshift, governed by the metric parameter $a$; the difference in the local
clock rates at the emission surface and the observer radius, also governed by the metric parameter $a$; and the reduction of the
neutrino flux, governed by the metric parameter $b^{-1}$.  All three of these effects are of roughly equal magnitude
and reduce the luminosities by a total factor of $a^2 b^{-1}$ evaluated at the $\nu$-sphere, as shown in Figure 3.  This figure
also shows the independent effects of $a$ and $b$ on the stationary-state neutrino transport.

The shock heating rate is proportional to the product of the luminosity and the square of the $\nu_{\mathrm{e}}$ rms energy. 
This means that any percentage change in these quantities add together.  Looking at Figures 2 and 3, we see a decrease of 8\% in
the $\nu_{\mathrm{e}}$ luminosity and a 3\% reduction in the  $\nu_{\mathrm{e}}$ rms energy.  These combine to give a 14\%
reduction in the heating rate.  Similar results are obtained for the $\bar{\nu}_{\mathrm{e}}$'s.

These are significant differences [\emph{e.g.}, see Burrows \& Goshy (1993), Janka \& M\"uller (1996), Mezzacappa \emph{et al.}
(1998), and Messer \emph{et al.} (1998)] and serve to illustrate the point that modeling core collapse supernovae without GR
hydrodynamics and transport leads to results that cannot be interpreted as realistic.  For more information on this subject, the
reader is referred to Bruenn, De~Nisco, \& Mezzacappa (1998).

\section{Nucleosynthesis}

r-process nucleosynthesis is believed to occur in a neutrino-driven wind emanating from the proto-neutron star after the
successful launch of the shock.  The r-process yields are a function of the $\nu_{\mathrm{e}}$ and $\bar{\nu}_{\mathrm{e}}$
luminosities and rms energies.  The luminosities affect the entropy, mass loss rate, and expansion time scale associated with the
wind, and the rms energies determine the neutronization of the wind. As an example of the impact of GR on the r-process,
consider its effect on the rms energies.  Because the $\bar{\nu}_{\mathrm{e}}$-sphere lies below the
$\nu_{\mathrm{e}}$-sphere, the $\bar{\nu}_{\mathrm{e}}$'s suffer a greater emergent redshift.  As can be seen in Figure 4, general
relativistic transport and hydrodynamics affects the ratio of the $\nu_{\mathrm{e}}$ and $\bar{\nu}_{\mathrm{e}}$ rms energies. 
This differential redshifting affects, in turn, the ratio of the number of $\nu_{\mathrm{e}}$'s to
$\bar{\nu}_{\mathrm{e}}$'s, at a given neutrino energy, and therefore, the neutronization of the wind.  This suggests that
general relativistic hydrodynamics and transport will be required to obtain accurate r-process yields.


\begin{references}

\item Bionta, R., \emph{et al.} 1987, \emph{Phys.~Rev.~Lett.}, \textbf{58}, 1494

\item Bruenn, S. W., De Nisco, K. R., \& Mezzacappa, A. 1998, in preparation

\item Burrows, A. \& Goshy, J. 1993, \emph{ApJ}, \textbf{416}, L75

\item Hirata, K., \emph{et al.} 1987, \emph{Phys.~Rev.~Lett.}, \textbf{58}, 1490

\item Janka, H.-Th. \& M\"uller, E. 1996, \emph{A \& A}, \textbf{306}, 167

\item Messer, O. E. B., Mezzacappa, A., Bruenn, S. W., \& Guidry, M. W., this volume

\item Mezzacappa, A., \emph{et al.} 1998, \emph{ApJ}, \textbf{495}, 911

\item Weaver, T. A. \& Woosley, S. E. 1998, in preparation

\item Woosley, S. E. \& Weaver, T. A. 1995, \emph{ApJ Suppl.}, \textbf{101}, 181

\end{references}
\end{document}